\documentclass[journal,comsoc]{IEEEtran}
\usepackage[T1]{fontenc}% optional T1 font encoding
\usepackage{amssymb}
\usepackage{epstopdf}
\usepackage{subcaption}
\usepackage{url}
\usepackage{xspace}
\usepackage{booktabs} 
\usepackage{color}

\ifCLASSINFOpdf
  \usepackage[pdftex]{graphicx}
\else
\fi

\usepackage{amsmath}
\begin{document}
\title{Multi-Dimensional Convergence \\ in Future 5G Networks}

\author{Marco~Ruffini~\IEEEmembership{Senior Member,~IEEE.}
% <-this % stops a space
% <-this % stops a space
\thanks{Financial support from Science Foundation Ireland (SFI) grant 14/IA/2527 (O'SHARE) and grant 13/RC/2077 (CONNECT) is gratefully acknowledged.}
\thanks{M. Ruffini is with the CONNECT research centre in The University of Dublin, Trinity College,
 e-mail: marco.ruffini@tcd.ie}}

\markboth{Journal of Lightwave Technology, 2016}%
{Shell \MakeLowercase{\textit{et al.}}: Bare Demo of IEEEtran.cls for IEEE Communications Society Journals}



\maketitle

\begin{abstract}
Future 5G services are characterised by unprecedented need for high rate, ubiquitous availability, ultra-low latency and high reliability. The fragmented network view that is widespread in current networks will not stand the challenge posed by next generations of users. A new vision is required, and this paper provides an insight on how network convergence and application-centric approaches will play a leading role towards enabling the 5G vision. 
The paper, after expressing the view on the need for an end-to-end approach to network design, brings the reader into a journey on the expected 5G network requirements and outlines some of the work currently carried out by main standardisation bodies. It then proposes the use of the concept of network convergence for providing the overall architectural framework to bring together all the different technologies within a unifying and coherent network ecosystem. The novel interpretation of multi-dimensional convergence we introduce leads us to the exploration of aspects of node consolidation and converged network architectures, delving into details of optical-wireless integration and future convergence of optical data centre and access-metro networks. We then discuss how ownership models enabling network sharing will be instrumental in realising the 5G vision. The paper concludes with final remarks on the role SDN will play in 5G and on the need for new business models that reflect the application-centric view of the network. Finally, we provide some insight on growing research areas in 5G networking.

\end{abstract}


% \fi
\IEEEpeerreviewmaketitle

\section{Introduction}
\IEEEPARstart{O}{ne} of the prevailing dilemmas operators are facing today is how to persuade residential users to take up faster broadband offers in areas where satisfactory speed (for example of the order of fifty to a hundred megabit per second) is already available at competitive price. Blame is typically given to the lack of marketable applications requiring speed of hundreds of megabit per second.
We argue this perception derives from an incorrect approach to the problem, and it is the cause of an un-converged vision of the network which is considered as a chain of separate sections only inter-connected at the IP protocol level.

In this paper we envision a unified view of the network that is achieved by scaling up network convergence to a point where the network can provide personalised connectivity to the application using it independently of the underlying technology, geographical location and infrastructure ownership.
While a number of research projects have emphasised the economic benefit of physical convergence of access and metro networks \cite{SARDANA-ECOC-2010},\cite{DISCUS-JOCN},\cite{COMBO-JOCN} we argue this new vision requires dealing with network convergence at a multi-dimensional level and focus on the following three dimensions: the spatial dimension dimension, the networking dimension and the ownership dimension \cite{OFC-tutorial-convergence}.
We also come to the conclusion that current business models based on broadband offering of peak bit rates does not have a place in the future, as the value for the end users lies in the correct delivery of the application, which should become the starting point of the value-chain. 

If we look at the past two decades we see that network convergence has driven the reduction in cost of ownership by moving voice services from the synchronous TDM transmission systems (e.g., Sonet and SDH) to the packet switched architecture used to transport Internet data, through the adoption of Voice over IP (VoIP) technology. This technological convergence also gave the operators the opportunity to offer bundled broadband services, such as triple play (e.g., voice, Internet and TV) and quadruple play (adding mobile phone to the mix), as a means to reduce their cost and to benefit from economies of scales associated to service consolidation.
This trend has evolved over the years, moving today its focus on the convergence of access and metro networks, which revolves around two key trends of infrastructure integration. The first is the consolidation of the number of central offices (COs), which is typically achieved by adopting fibre access architectures with longer reach, to bypass some of the current network nodes \cite{DISCUS-JOCN}. The second is the convergence of wireless and wireline networks, which typically focuses on the transport of data from mobile stations over shared optical access links.

This paper argues that while this access-metro infrastructure integration is a step in the right direction, it only represents part of the contribution that network convergence can provide in support of the application-centric network vision necessary to deliver future 5G services.

The paper is structured as follows. The next section provides an overview of the requirement of future 5G networks, based on the early work of industry fora. Section III gives a brief insight on the work carried out by some of the main standardisation bodies on 5G. While this short overview is far from being comprehensive, due to the large number of ongoing standardisation activities, it provides an insight on the main technologies being considered on wireless, optical and higher network layer technology.
After this, we investigate a number of research activities on converged network architectures that aim at providing an integrated framework to bring together different 5G technologies within a unifying and coherent network ecosystem. We report these activities under three distinct categories of convergence. The convergence in the space dimension, in section V, explores the use of long-reach access technology to enable central office consolidation. The convergence in the networking dimension, in section VI, discusses trends and options for end-to-end integration of different network types, focusing on fixed/mobile convergence and proposing a vision where agile data centres play a pivotal role in the virtualisation of network functions. The convergence in the ownership domain, in section VII, describes current work in the area of multi-tenancy for access network. Section VIII provides final remarks and discussions giving some insight on the role that SDN will play towards converged 5G networks and proposing the use of application-driven business models as a foundation of the 5G vision. Finally, section IX concludes the paper summarising the main key points and providing some insight on future research areas.

\section{Future 5G network requirements}
One of the main targets for operators in the design of next generation network architectures is to plan for an infrastructure that if 5G-ready, i.e., capable to support the next generation of applications and services. 
The Next Generation Mobile Networks (NGMN) Alliance was one of the first fora to come up with a set of use cases, business models, technology and architecture proposition for 5G \cite{NGMN-wp-2015}. The NGMN Alliance envisages the existence of three large application groups: 
\begin{enumerate}
\item Enhanced mobile broadband (eMBB), which aims at scaling up broadband capacity to deliver next generation of ultra high definition and fidelity video service with augmented reality. From a networking perspective the target is to provide every active user with a least 50 mbps everywhere, with enhanced targets of 300 Mbps in dense areas and up to 1 Gbps indoor.
\item Massive machine type communications (mMTC), aiming at scaling up the network to support tens of billions connected devices, with densities of up to 200,000 units per square km. The target is also to simplify the devices compared to 3G and 4G to enable ultra-low cost and low power consumption. mMTC is seen as a major enabler for the Internet of Things (IoT), which is the end-to-end ecosystem running on top of the machine-to-machine communication service.
\item Ultra reliable and low latency communication (uRLLC), aiming at decreasing end-to-end latency to below 1 ms and reliability levels above five nines. It is envisaged that such requirements will be crucial for applications such as automotive (e.g., inter-vehicle communication systems for accident avoidance) and medical (e.g., remote surgery), but could be extended to other types of tactile internet applications, which require ultra low latency user feedback mechanism.
\end{enumerate}

\begin{figure}[h]
    \centering
    \includegraphics[width=01\columnwidth]{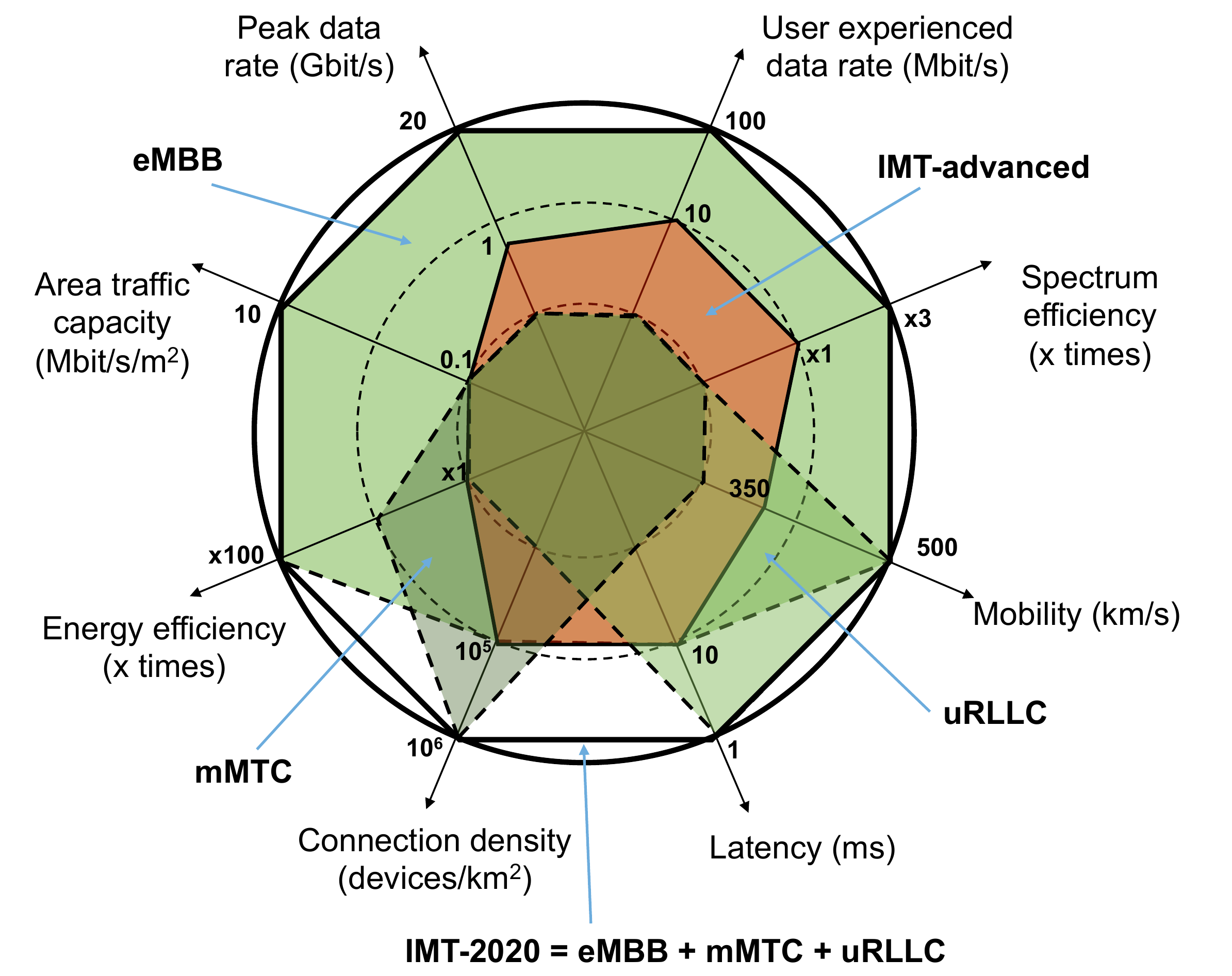}
    \caption{Key capabilities introduced by IMT-2020 networks, subdivided by application type, and compared to key capabilities enabled by IMT-advanced.}
    \label{fig:key-capabilities}
\end{figure}

Figure \ref{fig:key-capabilities} (derived form \cite{ITU-R-recommendation}) shows a summary of the increase in key capabilities of the network as the technology moves from IMT-Advanced \footnote{IMT-advances includes both LTE-Advanced and WiMAX release 2} (in red shade) towards IMT-2020 (in green shades). It should be noticed that although, collectively, IMT-2020 is expected to reach the most stringent requirements shown in the figure, no one application is expected to require them all simultaneously. Indeed the figure shows that such requirement can be further categorised according to the three application categories above. For example it is envisaged that eMBB will benefit from most of the enhancements, but it is not expected to require sub-ms latency (such as uRLLC applications) or density of devices above $10^5$ per $km^2$ (unlike mMTC applications requiring a density of up to $10^6$ per $km^2$). eMBB and eRLLC applications are in particular those that are expected to generate novel revenue streams for operators, especially from vertical markets \footnote{A vertical market is a market in which vendors offer goods and services specific to an industry, trade, profession, or other group of customers with specialised needs - Wikipedia definition}, as it is believed that 5G network infrastructures will enable the digitalisation of society and economy, leading to the fourth industrial revolution, impacting multiple sectors, especially the automotive, transportation, healthcare, energy, manufacturing and media and entertainment sectors \cite{5GPPP-verticals}.

Finally, an aspect of 5G networks of increasing importance is the ubiquity of broadband connection, as it is expected that the user experience continuity is not confined to urban districts but also extended to rural areas. The digital divide has become a major social and political issue worldwide, as fast broadband connectivity is now a commodity, and like drinkable water a right of every citizen. The European Commission has clearly stated broadband speed targets for the year 2020 of 100\% population coverage with at least 30Mbps, and 50\% with at least 100 Mbps \cite{EU-digital-agenda}. These requirements put additional pressure on developing access broadband architectures that are capable of reducing connectivity costs in rural areas, and are compliant with open-access models, which are necessary to operate in remote areas that require state intervention.

\subsection{Design for 5G}
While much of the 5G architecture and technologies are still undefined and under discussion, the diverse range of capabilities expected, often with conflicting goals, makes it clear that \textbf{\textit{5G is not simply the next-generation of 4G}}, or confined to the development of new radio interfaces, \textbf{\textit{but rather encompasses the development of an end-to-end system including multiple network domains, both fixed and mobile}}. Indeed we'll see in the next section that there are different technologies and standardisation bodies involved in the definition of the 5G ecosystem. 

In order to provide a level of insight on how to design a 5G-ready network, while not providing specific guidelines, the NGMN Alliance has attempted a definition of two general design principles \cite{NGMN-5G-prospects}:
\begin{itemize}
\item Provide expanded network capabilities: with the idea of pushing the network performance and capabilities over multiple directions, as exemplified in figure \ref{fig:key-capabilities}. 
\item Design intelligent "poly-morphic" systems: provide a malleable system that can be tailored to the application required. It is envisaged that virtualisation of network functions, control and data plane will play a pivotal role in enabling network flexibility.
\end{itemize}

The envisaged flexible network design implies flexibility in assigning network resources to applications, with diverse quality of service, reliability and availability requirements and will be reflected in updated business and charging models to drive the economics of the 5G ecosystem. A confirmation of this necessity is given by the  difficulty current operators  have to convince end users to buy ultra-high speed fibre access broadband where there are already other lower cost options offering satisfactory broadband speed. It is indeed increasingly difficult for end user to understand the practical benefit of faster access speeds, and justify its higher cost, when this does not assure the correct end-to-end delivery of services. In the current model the user pays for network connectivity, typically a flat rate, in addition to charges for the use of some applications, which however do not have the ability to view or influence the underlying end-to-end network performance. This model does not reflect the fact that the real value for the user is in the applications, which have diverse capacity and latency requirements, and whose \textbf{\textit{value per bit}} can vary substantially (for example if we compare a video on demand to a voice call or a medical device remotely sending body monitoring values to a medical centre).

Clearly this model needs to shift towards one that instead is linked to the application used, and makes the end user completely oblivious of the details of the underlying network performance. A practical, although basic, example of the implementation of this model is a recent deal where Netflix has agreed to pay Comcast for an improved delivery of its service over their network. 
Is the reliable service delivery through end-to-end quality of service (QoS) assurance, not the isolated increase in access data rate, that will generate new forms of revenue.
If the operators cannot deliver this type of service, it is likely that OTTs and industry verticals will continue to build their own dedicated network, which will lead to a fragmented suboptimal network development.

\subsection{Quality of service implementation}
Although the basic mechanism for QoS have been standardised many years ago through Integrated Services (IntServ) \cite{intserv} and Differentiated Services (DiffServ) \cite{diffserv}, attempts at using them in practice for end-to-end quality assurance have failed in the past. This was due to the complexity of implementing it across different domains at the granularity of the individual application and over-provisioning was used instead as a means to provide an average acceptable service. 

\begin{figure}[h]
    \centering
    \includegraphics[width=\columnwidth]{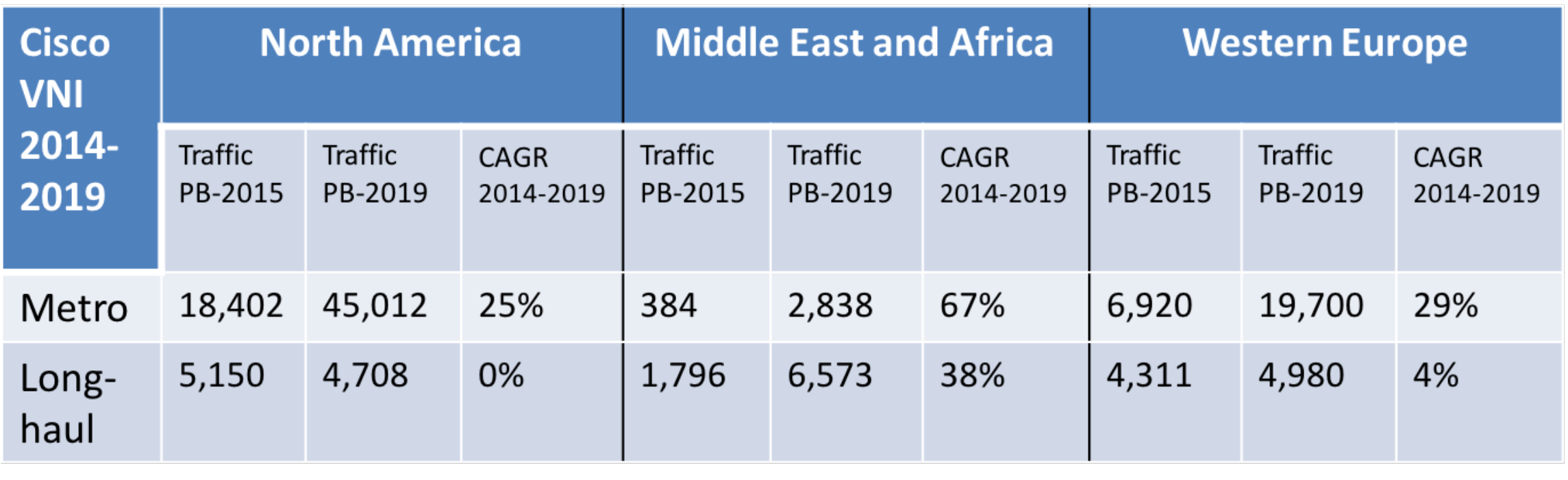}
    \caption{Metro vs Long-haul traffic growth estimates (Cisco VNI 2015) \cite{cisco-vni}}
    \label{fig:metro_lh}
\end{figure}

We argue that today the situation has changed for the following reasons and that QoS assurance will play a fundamental role in future networks.
Firstly, operators have now started to look for higher efficiency in their network as their profit have constantly declined, and massive overprovisioning ceases to be a valuable option. Secondly, the Internet architecture has migrated over the years from a hierarchical model, where end-to-end connectivity was provided by passing through multiple tiers, and crossing several network domains, to a model where many peering connections provide direct links between providers. This reduces substantially the number of domains crossed by the data flows which represented one of the main obstacles to QoS delivery. The reduction in average number of IP hops was also achieved through widespread installation of new data centres and the use of Content Delivery Networks (CDN) to reduce the distance between traffic source and destination. Indeed an analysis of current metro vs. long-haul traffic shows that the latter is in constant decrease (see Fig. \ref{fig:metro_lh}), suggesting that most end-to-end connections are confined within the metro area.
We argue that this change in circumstances will enable the success of large-scale end-to-end QoS delivery. This will be facilitated by the development of open and programmable networks, based on the Software Defined Networking (SDN) paradigm, which will automate most of the QoS configuration hiding its complexity to network administrators. Indeed many operators are already investigating the use of SDN in their network, with some making it their main short-term goal \cite{ATT-domain}, and discussions have already started among industry fora for delivery of Broadband Assured Services \cite{BBF-SD-377}.

\section{Standardisation activities towards \\ the 5G vision}
\begin{figure}[h]
    \centering
    \includegraphics[width=\columnwidth]{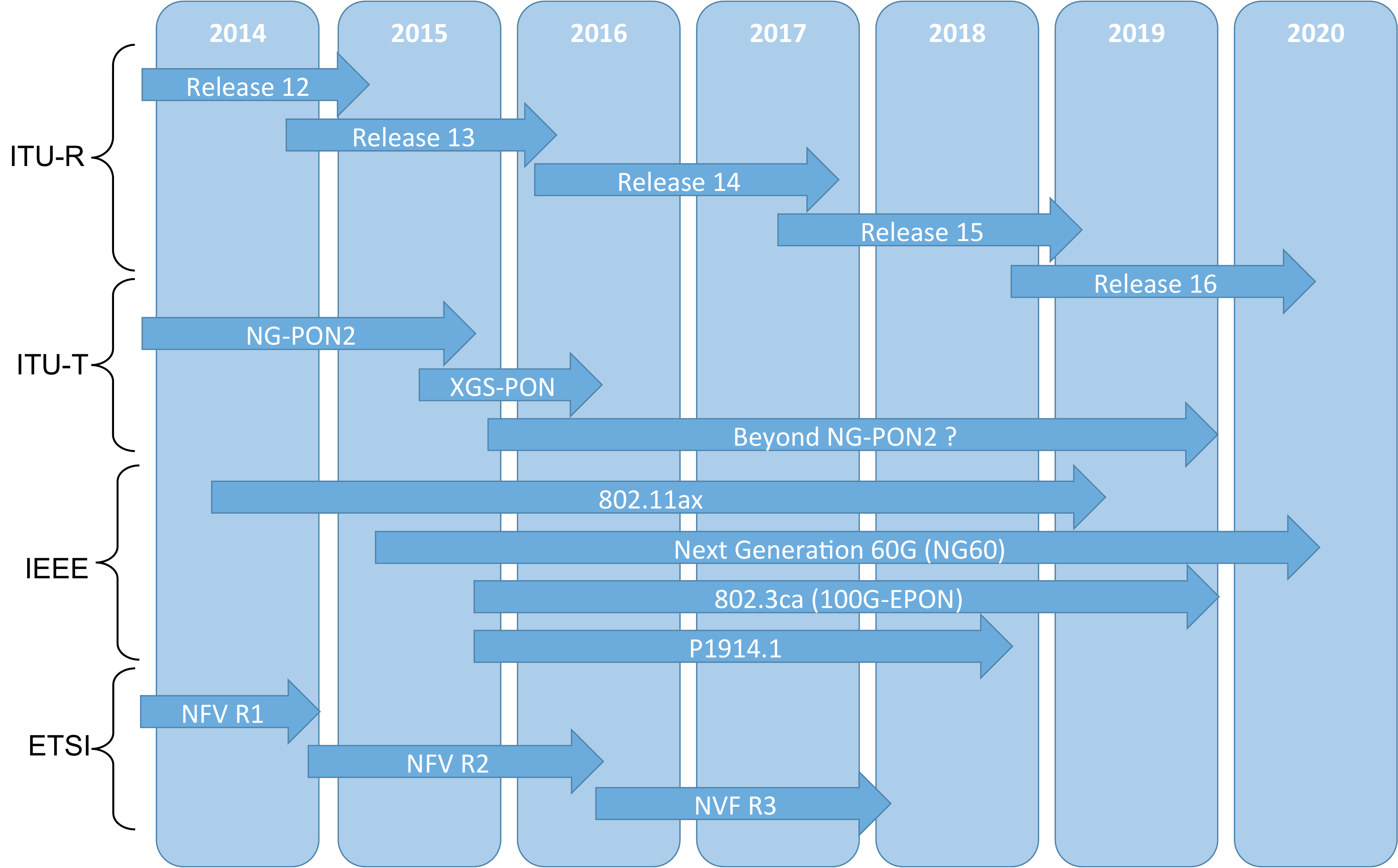}
    \caption{Roadmap of some of the major standardisation bodies}
    \label{fig:standard_roadmap}
\end{figure}

While the requirements and use cases are still work in progress and are progressively refined as discussions on 5G carry on, standardisation bodies have started building up a roadmap to evolve the current technology towards 5G. This section reports on some of the most relevant standardisation activities, considering both fixed and mobile networks, that contribute at different levels towards the realisation of the overall 5G vision.
Here we report some highlights of the standardisation roadmap for major bodies like the ITU, IEEE, ETSI and ONF, which are summarised in figure \ref{fig:standard_roadmap}.

\subsection{ITU}
Having standardised most of the previous generations of mobile communications, one of the most active bodies in the standardisation of radio aspects of 5G is the ITU Radiocommunication Sector (ITU-R), through their International Mobile Telecommunication \footnote{This is the programme that has defined first, second, third and fourth generations of mobile networks} IMT-2020 programme. Releases 13 and 14 (recognised as part of the LTE-Advanced \cite{LTE-A} framework) are setting up the basis towards the required 5G enhancements, considering technologies such as Full-Dimension Multiple Input Multiple Output systems (FD-MIMO) with 2D arrays of up to 64 antennas, Licensed-Assisted Access (LAA) enabling the joint usage of unlicensed and licensed spectrum, and enhanced carried aggregation to increase the number of carries from 5 to 32 \cite{comms-standard-rel13-14}. However it is expected that the full IMT-2020 specification will be provided with release 16 (with an expected release date around the year 2020), where new air interface for operations above 6GHz will be developed in addition to evolutions that are backward compatible with LTE (i.e. on radio frequencies below 6GHz). IMT-2020 is also targeting the Vehicle-to-X (V2X) type of use cases, focusing specifically on Vehicle-to-Vehicle (V2V), Vehicle-to-Pedestrian (V2P) and Vehicle-to-Infrastructure/Network (V2I/N) \cite{5gPPP-automotive-white-paper}. These efforts include studies on enhancements to resource allocation mechanisms,  to improve robustness, latency, overhead and capacity of the mobile communication system.

From an optical access network perspective, while not explicitly considered a 5G evolution, ITU-T has recently standardised the XGS-PON \cite{XGS-PON}, for cost-effective delivery of symmetric 10 Gbps to the residential market in the shorter term, and NG-PON2 \cite{NG-PON2}, offering up to 80 Gbps symmetric rates over 8 10G wavelength channels. The ITU optical access group is currently investigating what other technologies to consider beyond NG-PON2, with proposals spanning from amplified Optical Distribution Network (ODN) for longer reach to coherent transmission optical access. Considering yearly traffic growth rates of about 30\% (for most developed countries) \cite{cisco-vni}, it is expected that NG-PON2, although only exploiting a small portion of the available optical spectrum, will be able to deliver the capacity that residential users might need for the foreseeable future, and that future generation of standards should focus on aspects of cost reduction and network flexibility, for example to support mobile transport, business and residential types of services in the same PON infrastructure \cite{Nesset-OFC-2016}. 

\subsection{IEEE}
Another body considering 5G adaptations to its standards is the IEEE, with its 802.11 group working on two main activities. The 802.11ax, labeled High Efficiency Wi-Fi, enhances the 802.11ac by aiming at throughputs of the order of the 10Gb/s per access point, through a combination of the MIMO and Orthogonal Frequency Division Multiplexing (OFDM) technologies and working on frequency bands between 1 and 6 GHz. The Next Generation 60GHz (NG60) aims instead at evolving the 802.11ad air interface to support data rates above 30 Gb/s. The high frequency limits its operation to line of sight transmission, targeting applications such as wireless cable replacement, wireless backhaul and indoor short-distance communications. 

From an optical access perspective, IEEE is also updating the 10Gb/s PON interface with the new 802.3ca standardisation effort, investigating possible physical layer data rates of 25, 50 and 100 Gb/s, expected to be fully released by the end of 2019. It is also worth mentioning the IEEE effort on Standard for Packet-based Fronthaul Transport Networks (P1914.1). This provides a practical solution to the fixed-mobile convergence problem, aiming to deliver an Ethernet-based transport, switching and aggregation system to deliver fronthaul services.

\subsection{ETSI}
In addition to physical layer standardisation, the 5G ecosystem requires standardisation also of higher layers. In particular Network Function Virtualisation (NFV) \cite{NFV} is identified as a target for 5G networks \cite{Comms-standard-virtualisation-5G}, to increase network programmability and reduce cost of ownership by moving network functions from proprietary hardware to software running on general purpose servers (a concept also known as "softwarisation"). ETSI has been active on NFV from its emergence, publishing the first release of its specification in December 2014, where it provided an infrastructure overview, an architectural framework, and descriptions of the compute, hypervisor and network domains of the infrastructure. The architecture has three main constituent elements \cite{ETSI_whitep-2}: the Network Function Virtualisation Infrastructure (NFVI), which provides the commodity hardware, additional accelerator components and the software layer that abstracts and virtualises the underlying hardware; the Virtualised Network Function (VNF), which is the software implementation of the required network function and runs on top of the NFVI; and the NFV Management and Orchestration entity (M\&O) taking care of the lifecycle management of the physical and software resources and providing an interface to external Operartion Support Systems (OSS) and Business Support Systems (BSS) for integration with existing network management frameworks. The second NVF release, covering the working period from November 2014 to mid 2016, provided NFVI hypervisor requirements, functional requirement of management and orchestration, hardware and software acceleration mechanism for the virtual interfaces and virtual switch. It also expanded the architectural framework with the further specification of the Management and Orchestration entity (dubbed MANO) with the definition of \cite{ETSI_whitep-3}: the Virtualised Infrastructure Managers (VIM), which performs orchestration and management functions of NFVI resources within a domain; the NFV Orchestrator (NFVO), performing orchestration of NFVI resources across multiple domains; and the VNF Manager (VNFM) carrying out orchestration and management functions of the VNFs.
ETSI's work towards the third NFV release has only recently started and is expected to target topics such as charging, billing and accounting, policy management, VNF lifecycle management and more.

\subsection{ONF}
The Open Networking Forum is the reference consortium for the standardisation of the Software Defined Networking paradigm. After the release of the OpenFlow (OF) v1.0 in December 2009, their work has progressed to define a plethora of updates, with new releases every few months (for this reason the ONF roadmap is not reported in figure \ref{fig:standard_roadmap}). While the OF v1.0 specification has evolved to v1.3.5, other two releases have progressed in parallel to allow the development of more unconventional versions of the protocol, that was not seen as essential for all vendors. For example v1.4 allowed for a more extensible protocol definition and introduced optical port properties. Version 1.5 introduced the concept of Layer 4 to Layer 7 processing through deep header parsing and the use of egress tables to allow processing to be done in the context of the output port.
While the OF specification targets the control plane functions, the management and configuration specification where carried out through the 'Of-Config' protocol releases, currently at version v.1.2.
Additional specification were also released to target transport networks, covering aspects of multi-layer and multi-domain SDN control, together with many technical recommendations on SDN architectural aspects.

\section{Enabling the 5G vision through network convergence}
The previous two sections have provided an insight on expected 5G network requirements and given a brief description of the roadmap pursued by some of the most relevant standardisation bodies. 
A first set of activities, at the lower network layers, focus on the enhancement of network performance, addressing higher cell density, higher peak rate and energy efficiency, as well as scalability, latency reduction and higher reliability. This is reflected for example by the work carried out by ITU-R, ITU-T, and IEEE cited above. 
A second set of activities, at the higher network layers, are targeting software-driven approaches to resource virtualisation and control, through network virtualisation, NFV and SDN control layers, which is strongly driven by ONF and ETSI. This is believed to be a distinctive feature of 5G networks, to satisfy the 'Poly-morphic' design principle and provide the flexibility and automation required to accommodate the envisaged diversity of requirements.
In addition, it is envisaged that a third set of activities should focus on new business and network ownership models that will have to emerge to make the integration of all the various components profitable for the market players. The NGMN Alliance for example recognises that the creation of a valid and all-encompassing business case for 5G is pivotal for the sustainability of the entire 5G ecosystem \cite{NGMN-5G-prospects}.

\begin{figure}[h]
    \centering
    \includegraphics[width=\columnwidth]{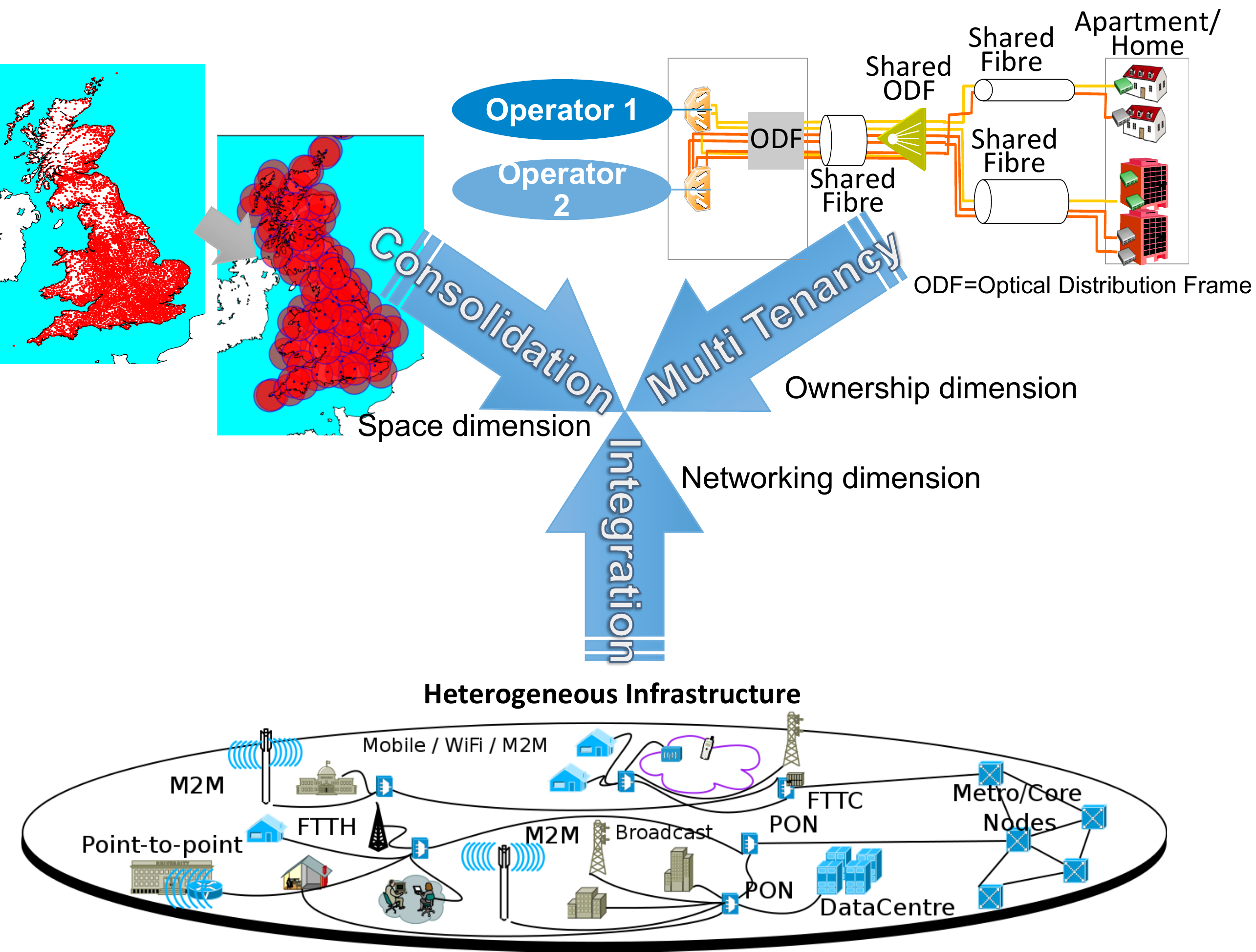}
    \caption{Introducing the concept of multi-dimensional network convergence}
    \label{fig:convergence_all}
\end{figure}

While standards are taking care of some of the potential emerging technologies they only operate within specific technologies and typically do not provide the end-to-end view that is required for the 5G vision. In this tutorial we complement the analysis on requirement and standardisation activities on 5G with an overview of a number of recent research projects and activities on network architectures aiming towards a unified end-to-end view of the network. 
We propose to categorise such activities across three complementary dimensions of network convergence, summarised in figure \ref{fig:convergence_all}, which can contribute the overall architectural framework to bring together different 5G technologies within a unifying and coherent network ecosystem.
The convergence in the space domain signifies the consolidation of many of the current network nodes: the the top left of figure  \ref{fig:convergence_all} shows a node consolidation analysis for the UK that could reduce the number of total nodes from over 5,000 to less than 100 \cite{DISCUS-JOCN}; the convergence in the networking dimension, at the bottom of the figure, designates the integration of heterogeneous infrastructure across different network segments to allow end-to-end control of networking resources; the convergence in the ownership domain, in the top right hand side of the figure, denotes the concept of multi-tenancy across network resources, allowing multiple operators to share physical network infrastructure \cite{Cornaglia-OFT}.
Overall, while the network can benefit from all three types of convergence, some of them have contrasting requirements, for example considering the trade-off between node consolidation, requiring longer links with higher latency, and support for some of next generation mobile services, which present today very tight latency constraints.  We anticipate that some of these issues have not yet been resolved and should be addressed in future research projects.
The next three sections of this paper provide an in-depth analysis of these network convergence topics. At the end of each section we summarise how the research activities described contribute towards the realisation of the 5G vision, and provide an overall outline in figure \ref{fig:req_mapping}.

\
\

\section{Convergence in the space dimension: \\ node consolidation}
The first category of the convergence we discuss is that in the space dimension, which is achieved by integrating the physical architecture of access and metro networks to enable node consolidation, i.e. a significant reduction in the number of central offices. The aim of this convergence is manifold, providing capital and operational cost saving through a massive reduction in the number of electronic port terminations (i.e., optical-electrical-optical (OEO) interfaces), in addition to simplification of network architecture and management.
The Long-Reach Passive optical network architecture, originally introduced in \cite{LR-PON-BT-tech}, and further developed by the European DISCUS consortium \cite{DISCUS-JOCN}, targets exactly this idea. By extending the maximum PON reach to distances above 100 km through the use of in line optical amplifiers in the optical distribution network, LR-PON can transparently bypass the current metro transmission network, linking directly the access fibre to a small number of Metro-Core (MC) nodes, which serve both network access and core. 

\begin{figure}
    \centering
    \begin{subfigure}[b]{0.4\textwidth}
        \includegraphics[width=\textwidth]{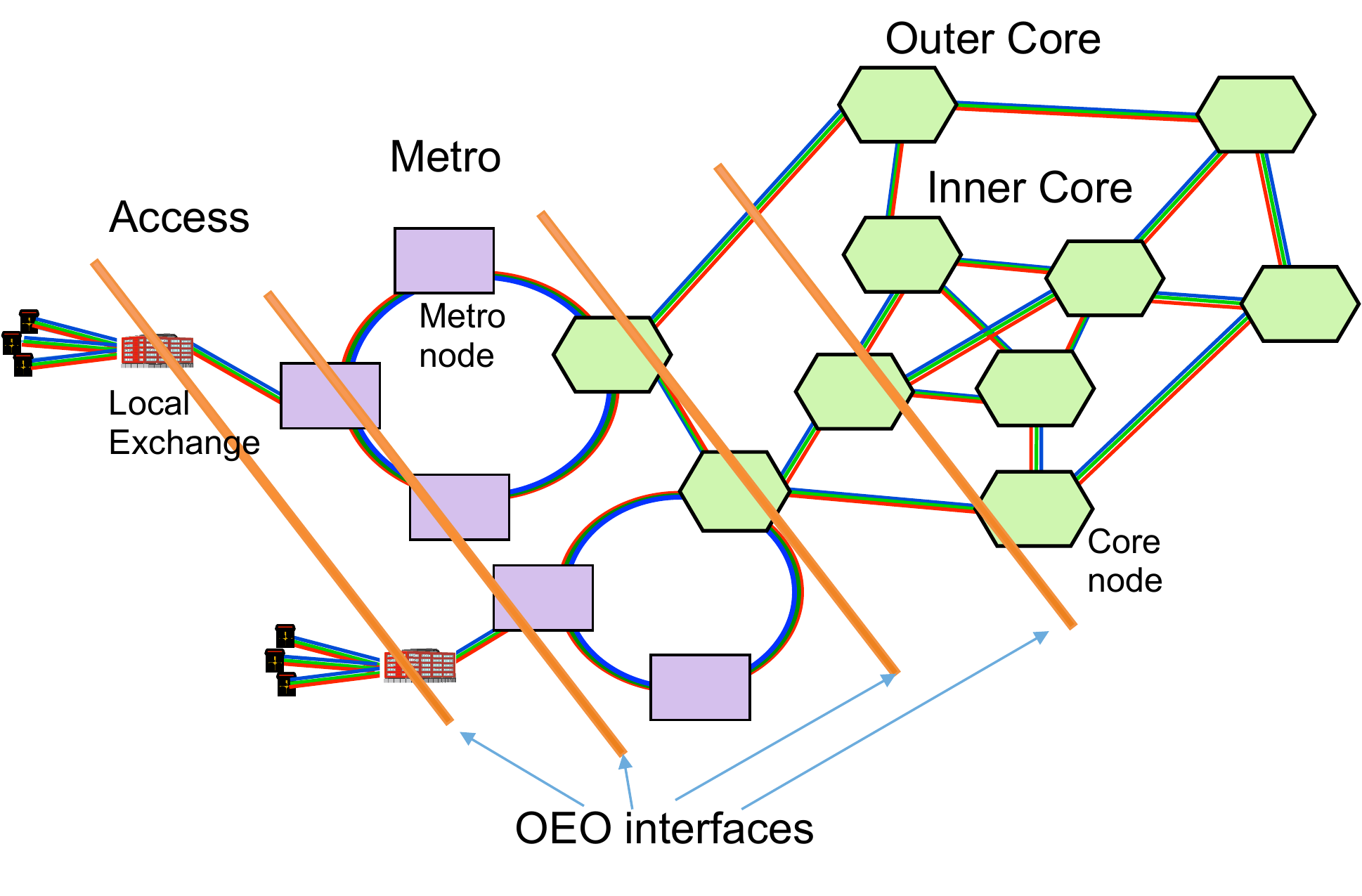}
        \caption{Traditional network architecture with separate access, metro and core}
        \label{fig:arch_today}
    \end{subfigure}
    ~ %add desired spacing between images, e. g. ~, \quad, \qquad, \hfill etc. 
    \begin{subfigure}[b]{0.4\textwidth}
        \includegraphics[width=\textwidth]{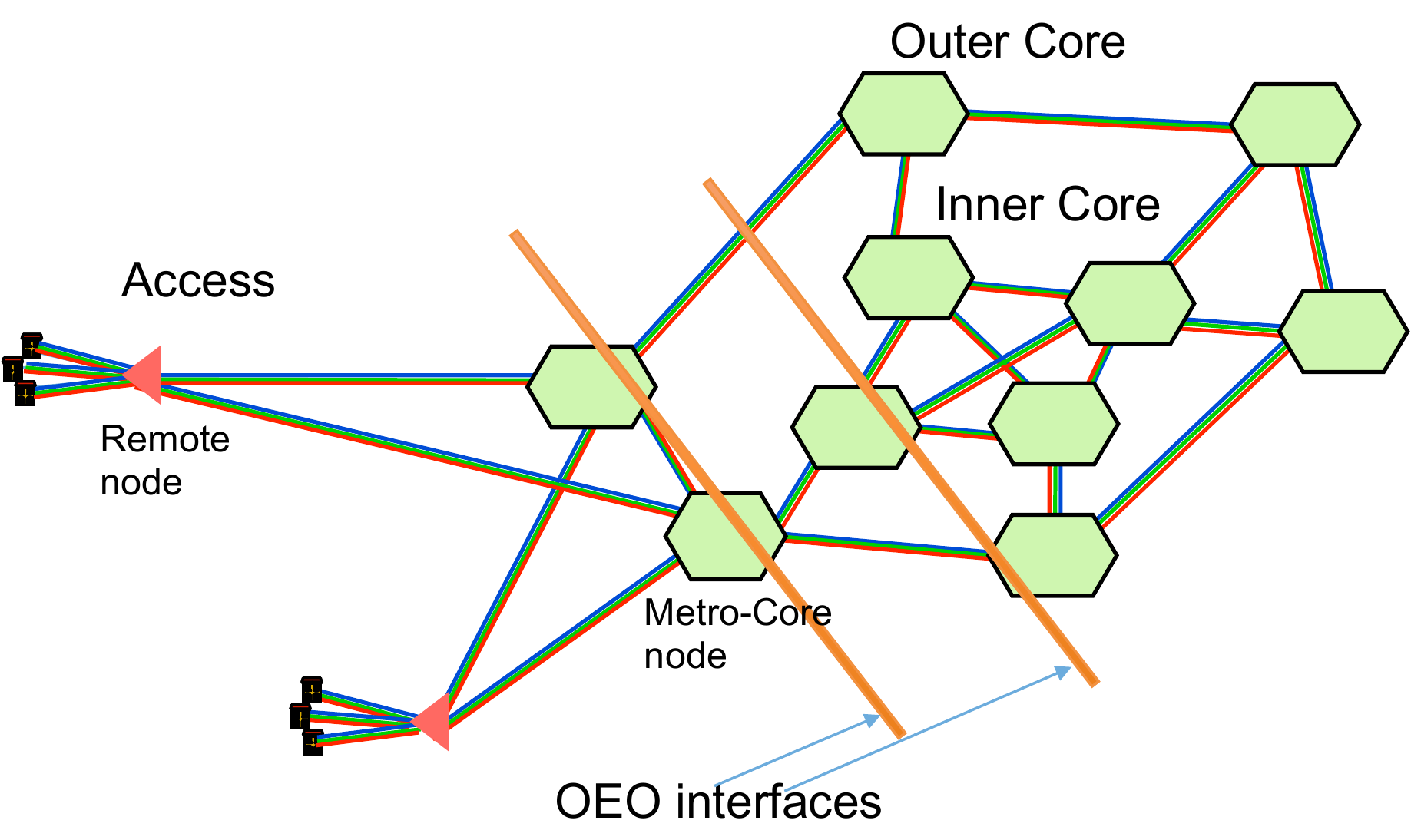}
        \caption{Long-Reach PON architecture with consolidation of access and metro networks}
        \label{fig:arch_LR}
    \end{subfigure}
    ~ %add desired spacing between images, e. g. ~, \quad, \qquad, \hfill etc. 
    \begin{subfigure}[b]{0.4\textwidth}
        \includegraphics[width=\textwidth]{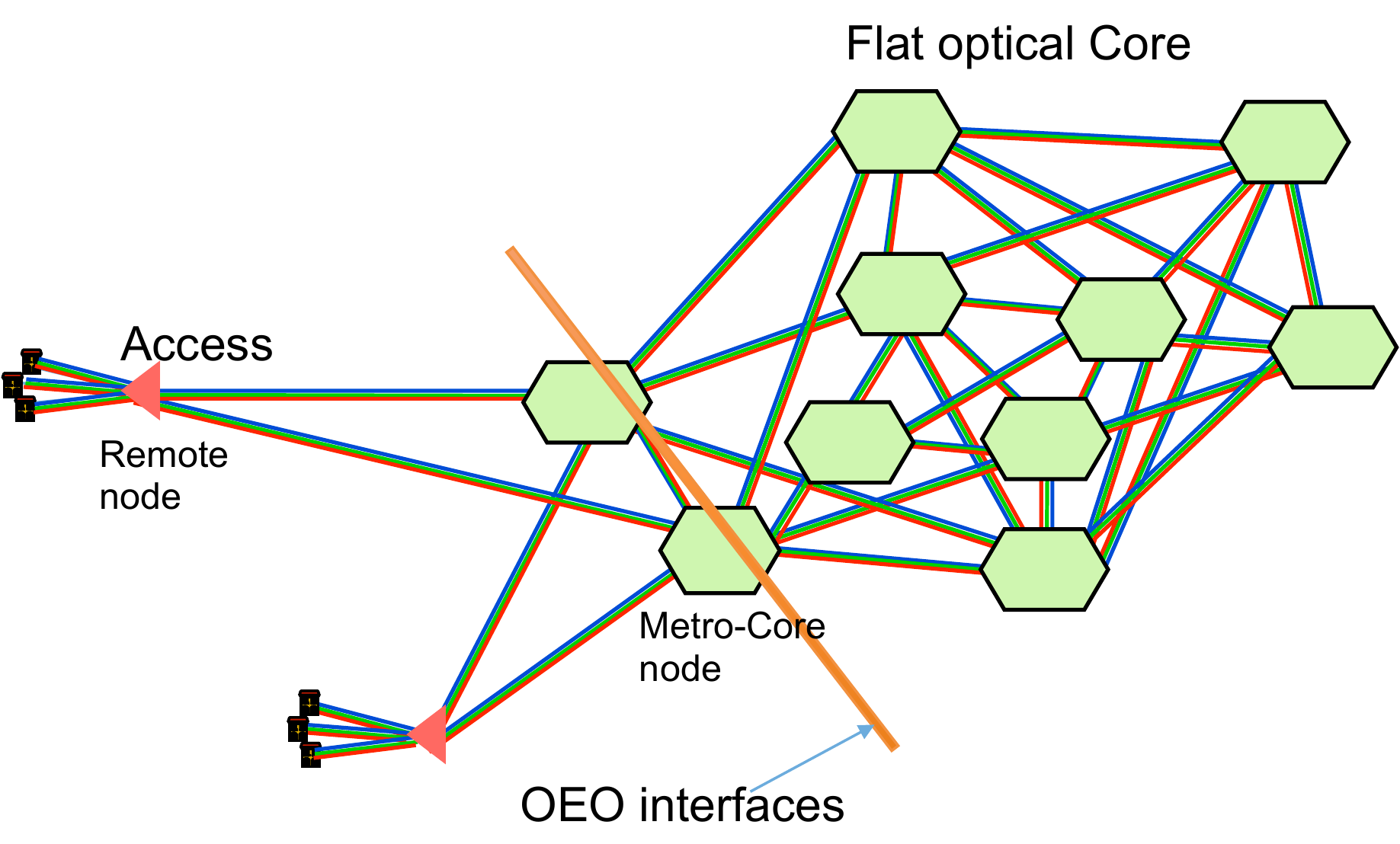}
        \caption{Long-Reach PON architecture with full flat optical core}
        \label{fig:arch_LR_flat}
    \end{subfigure}
    \caption{Evolution of the network architecture from traditional separation of access, metro and core to Long-Reach PON with flat optical core}
    \label{fig:arch_evolution}
\end{figure}
This evolution is shown in figure \ref{fig:arch_evolution}, where starting from the traditional architecture that separates access, metro and core in figure \ref{fig:arch_today}, the access-metro convergence operated by LR-PON can remove at least two levels of OEO interfaces in the network leading to the scenario in figure \ref{fig:arch_LR}. Studies carried out for a number of European countries \cite{JOCN_prot}, \cite{DISCUS_D76}, show that typically the number of central offices can be reduced by two orders of magnitude.
An additional benefit of LR-PON networks is that by bringing the total number of nodes (e.g., within a national network) below a given threshold (typically in the order of a hundred), it enables a fully flat core network, moving to the architecture represented in figure \ref{fig:arch_LR_flat}. The flat core interconnects the metro-core nodes through a full mesh of wavelength channels, using transparent optical switching to bypass intermediate nodes, thus eliminating the inner core OEO interface. With this architecture the only OEO interface is in the MC node, as packets are only processed electronically at the source and destination nodes. Studies carried out in \cite{ONDM_flatcore} have shown that for values of sustained user traffic above a certain threshold (about 7Mb/s for their scenario) the flat core becomes more economical than the current hierarchical model based on outer and inner cores. The large reduction of OEO ports can achieve large capital and operational saving, as shown in \cite{Payne_2009}, and a 10 times reduction in overall power consumption \cite{DISCUS_D28} compared to "Business as Usual" scenarios.

It should be stressed that this architecture is capable of achieving such advantages because it considers the network from a \textbf{converged, end-to-end perspective}, rather than optimising separately access, metro and core. For example, one of the  crucial parameters in long-reach access is the optimal value for the maximum optical reach, which cannot be determined if the access-metro part is considered independently from the core. The end-to-end perspective, which brings the core into the picture, provides a convincing argument for identifying such parameter, which is the value required to reduce the number of MC nodes to a level that enables their interconnection through a full mesh of wavelengths, (i.e., a flat core). Under such circumstances it was shown in \cite{Payne_2009} that the deployment of the Long-Reach architecture can be even more cost effective than fibre to the cabinet, as the cost reduction in the core network can subsidise the cost of deploying fibre infrastructure in the access. 
In addition the Long-Reach access is beneficial for lowering cost of fibre deployment in sparse rural areas, a primary target for every government that wants to reduce their extent of financial  intervention.

The studies mentioned above on the benefits of implementing node consolidation reveal that it can contribute to improving the business case for 5G, by reducing the overall network cost and energy consumption, thus providing the foundation for an architecture capable to deliver broadband fibre connection to a larger number of users, creating a ubiquitous optical access network ecosystem. The envisaged high-capacity and dynamic multi-wavelength PON architecture also allows flexibility in the type of service that can be offered, as each user end point can avail, on-demand, of different types of services (e.g., from dedicated point-to-point to shared wavelength). 
However, while this architecture can provide a basis towards the design of future networks, it lacks flexibility in the Optical Distribution Network side, as every user signal needs to be sent to the main central office for further processing. 
Based on this idea, in \cite{Pfeiffer-OFT} the architectural concepts were further developed to provide additional flexibility with the inclusion of Branching Nodes (BN) positioned in place of some of the remote nodes in the LR-PON architecture, providing wavelength routing and signal monitoring, in addition to optical amplification. This for example would allow the add/drop of optical signals close to the end point for local processing, when and where required, thus significantly reducing transmission distance to satisfy the requirements of those applications with strict latency constraints. The addition of a flexible BN in the network architecture further contributes to the 5G poly-morphic design principle. 

 \section{Convergence in the networking dimension}

The second category of convergence we discuss is in the networking dimension, aiming towards the integration of different types of networks, mobile, fixed and data centres, to enable end-to-end resource management. Among the three convergence dimensions, this is the one that most impacts the 5G vision, as it involves integrated development with the wireless access.

\begin{figure*}
    \centering
    \includegraphics[width=0.8\textwidth]{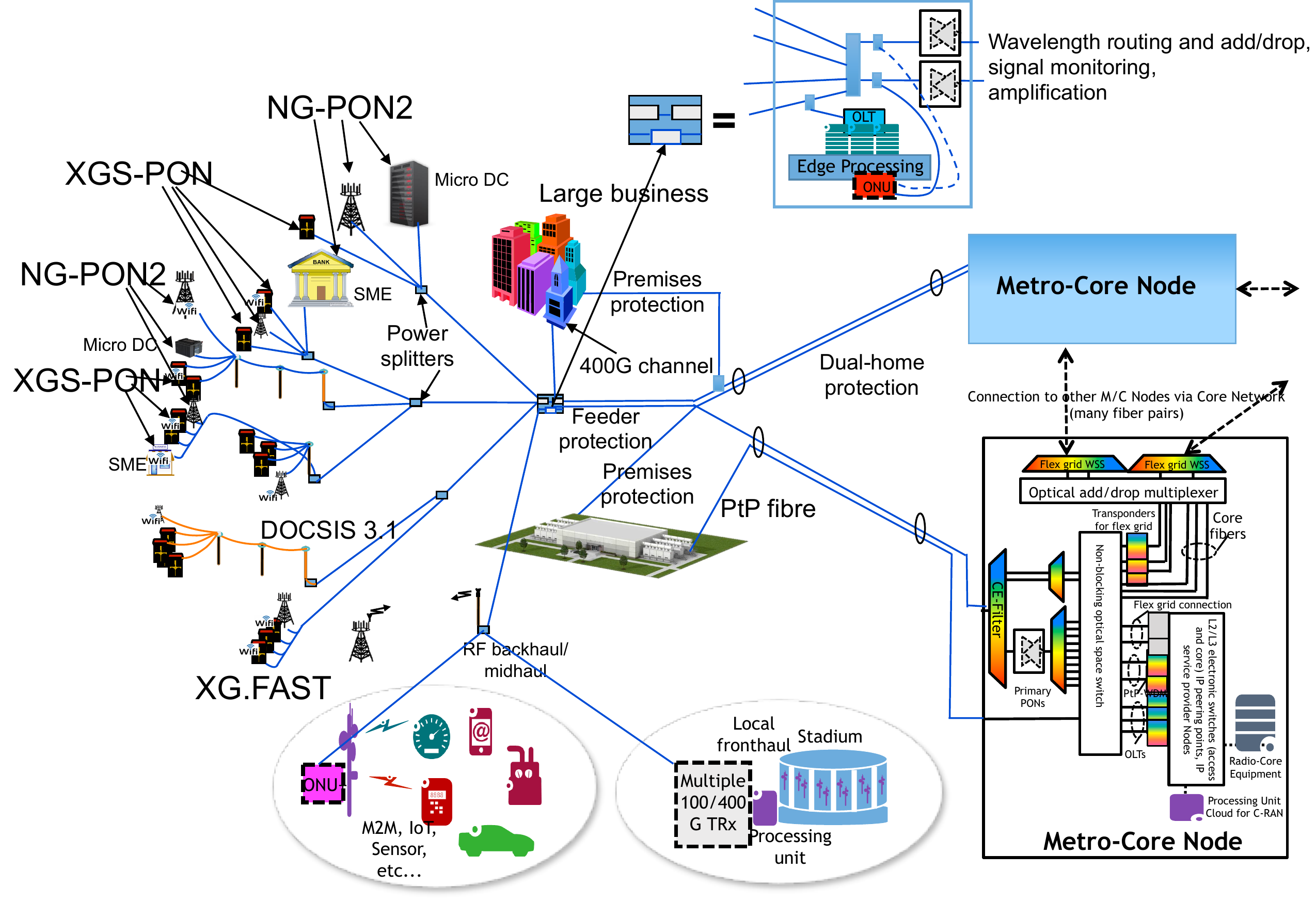}
    \caption{A converged access-metro architecture}
    \label{fig:pon_multi-service}
\end{figure*}

 As mentioned in the introduction, an example of large-scale network convergence was the migration of voice services from circuit-switched to packet switched IP networks, generating substantial savings in cost of network ownership for operators. Extending the concept to the access network, a strong integration between mobile and fixed technologies within a ubiquitous fibre deployment provides the resources to serve a wide range of user and services for several years to come, as any point of access can in principle deliver several terabits per second of capacity and bring such capacity from one end of the network to the other. The idea is that to build a flexible network architecture where multiple technologies can converge and provide a pool of diverse resources that can be virtualised, sliced and managed to provide the required end-to-end connectivity to applications. An example of the converged architecture vision is shown in figure \ref{fig:pon_multi-service}.

In addition, configurability and openness to upcoming technologies is a key factor when considering the unpredictability of technology adoption and financial return on investment. The current debate on GPON upgrade paths is an example. It was previously believed that most operators would skip the XG-PON standard (defining a single channel operating at 10Gbps downstream and 2.5Gbps upstream) \cite{XG-PON} and transition from GPON directly to NG-PON2. However, due to the high initial cost of NG-PON2 equipment and uncertainty on the additional revenue that could be generated, it is likely that residential users will be upgraded to XGS-PON (one single channel operating with symmetric 10G rate), while, simultaneously, NG-PON2 will be adopted for offering enhanced services and higher flexibility for business users and mobile cell interconnection (e.g., thought standard backhaul, fronthaul or midhaul \cite{Pfeiffer-JOCN}).

Indeed it is envisaged that such higher flexibility will be required in next generation 5G networks, where a ubiquitous and flexible fibre access network will provide connectivity to services with different capacity, service type and reliability requirements to different end points. For example, respectively, offering capacity of tens of Mb/s to hundreds of Gb/s, with quality ranging from best effort to assured capacity, with reliability spanning from no protection to dedicated 1+1 end-to-end protection, to diverse type of customers such as residential users, mobile stations, business and governmental organisation, data centres and local caches, etc.

Finally, among the available technologies in the access, we should consider that copper will still play an important role in the near future, as access fibre deployment is usually characterised by a long return on investment, leading some operators to use new high-speed copper access technology as intermediate steps towards FTTP. For example G.FAST \cite{G.FAST} is currently being considered to deliver few hundreds Mbps over distances up to about 300 m. In many cases G.FAST will make use of a PON as backhaul, but use the existing copper pair for the last drop. Even faster rates, up to 10Gbps, can be achieved over copper with XG.FAST technology \cite{XG.FAST}, which can terminate the XGS-PON or NG-PON2 fibre and use the existing twisted copper pair over few tens of meters to reach the household. This could be used for example in cases where the fibre deployment over the very last tens of meters of the drop are particularly expensive or inconvenient. Other users might be connected through a cable system with the new DOCSIS 3.1 standard capable of reaching 10Gbps downstream: these can already be backhauled through GPON \cite{RFoG}, while new proposals are currently being developed to guarantee compatibility with NG-PON2 \cite{PDB}. 

\subsection{Fixed/mobile convergence}
One interesting scenario of network convergence is the use of PONs to provide connectivity to mobile base stations. Wireless network capacity has increased over the past 45 years by one million times, a trend also known as Cooper's law of spectral efficiency. If we look at how this was accomplished we see that a factor 25 increase was enabled by higher efficiency of transmission technology, an additional factor of 25 by the use of more spectrum, but the vast majority, counting for a factor of 1600 was given by densification of cells, i.e. enabling spacial reuse of frequency channels. Looking at the next five years, a major goal of some of the network vendors working towards 5G is to provide a further capacity increase of up to a thousand times \cite{nokia_WP}. Although it is still uncertain whether this is achievable, any such increase in capacity is likely to follow a similar split between improvement of transmission technology, spectrum resources and densification. 
Beside the challenges with delivering such a high density of capacity on the radio interface (e.g., due to interference and frequency reuse issues) there is a comparable challenge for linking an ever increasing number of small cells to the rest of the network in a way that is cost effective. A solution to this problem that is gaining traction is to use PON networks: even though PONs were initially deployed as a means to bring ultra-fast broadband to residential users, the high capacity the fibre provides and the high degree of reconfigurability enabled by the simple power split architecture makes it an ideal candidate also for connecting base stations.

While with 3G systems, base stations were typically connected to the network through backhauling, i.e., at layer 2 or 3, massive densification of next generation mobile networks brings new challenges that cannot be solved through simple backhauling. Deploying a large number of cells can be prohibitively expensive when accounting costs for the base stations and the rental of the space where all its processing equipment is located. A solution that is becoming increasingly popular is to use fronthaul. The idea stems from the extension of a transmission protocol called Common Public Radio Interface (CPRI) that was used to link the antenna at the top of a base station mast with the BaseBand processing Unit (BBU) equipment at ground level though optical fibre. 
This eliminated the need for the RadioFrequency (RF) interface between antenna and BBU, thus reducing complexity, space requirements, heat dissipation and ultimately costs.
The concept was extended to BBU hoteling, where the link between the antenna mast and the BBU is increased to a few kilometres to place BBUs from different masts into one building in order to reduce deployment costs and enhance security, reliability and ease of management. 

It should be stressed however that while adopting smaller cells has the benefit of offering higher data rates per user as the total cell capacity is shared among a smaller number of users, it also reduces cell utilisation as the lower number of connected users reduces the statistical multiplexing gains compared to larger cells. Thus the need for additional savings has lead  to the concept of BBU pooling, where multiple radio heads (RRH) are multiplexed into a smaller number of BBUs. Besides improving BBU utilisation, by operating statistical multiplexing also within the BBU units, this mechanism allows centralised control of multiple RRHs, enabling the use of advanced LTE-A techniques such as Coordinated Multi-Point (CoMP), coordinated beamforming, and Inter-Cell Interference Cancellation (ICIC).

Finally, the third step of this convergence is to run the BBU as software on a virtual machine (software implementation of the LTE stack are already commercially available \cite{amarisoft}), in public data centres, a concept known as Could Radio Access Networks (C-RAN). This is fully in line with the 5G vision of NFV.

From a technical perspective, the issue with fronthauling is that it operates by transmitting I/Q samples, which increases the transmission rate over fibre by a factor of 30 \cite{Pfeiffer-JOCN}, compared to backhauling. In addition, since sampling occurs whether or not an actual signal transmission is in place over the radio interface, this rate is fixed, independently of the amount of data used in uplink or downlink by mobile users. Thus a large cell providing an aggregate rate of about 10Gbps, for example using an 8x8 MIMO array, 3 sectors and 5 x 20MHz channels, will require an approximate fronthaul transmission rate of about 150Gbps, while a small cell with simpler 2x2 MIMO, 1 sector on a single 20MHz channel, would still require 2.5Gb/s, while offering a radio data rate below 150Mbps. 

In addition, fronthaul imposes an upper bound to the latency between the remote radio head and the BBU, due to a maximum Round Trip Time (RTT) of 3 ms between handset and the Hybrid Automatic Repeat reQuest (HARQ) processing block in the BBU. Considering the latency introduced by the different mobile system processing blocks, typically a budget of up to $400 \mu s$ is allowed for the optical transmission system, imposing a maximum distance between RRH and BBU of about 40km.

Data compression techniques have been proposed in the literature \cite{ETSI-ORI} in order to reduce the large capacity requirements of fronthaul. Another concept, known as mid-haul, originally introduced in \cite{BellLabs-tech-j}, that is becoming increasingly popular is moving the physical split between RRH and BBU to a different point of the LTE stack. This can reduce the optical transmission rate by a factor of 5 \cite{Pfeiffer-JOCN} to 10 \cite{Miyamoto-split-PHY-OFC} compared to fronthaul and restore the dependency from the mobile data traffic volume, thus bringing back the possibility of operating statistical multiplexing of a number of mid-haul transmission systems. However there is a trade-off with some of the LTE-A functionalities as for example CoMP performs better for solutions placing the split closer to the RRH \cite{Kani-OFC}.

For all cases above, latency remains an open issue: since the HARQ processing block is located in the MAC, a solution that would position the HARQ in the RRH would be almost equivalent to backhauling.

A number of research projects have worked to provide architectural solutions to the fixed-mobile convergence issue. The FP7 COMBO project has proposed a number of solutions based on PONs to bridge the capacity gap between the base station and the central office in cost effective manners. The idea is to use different technologies in the ODN: shared TWDM PON channels can be used to multiplex backhaul signals from base stations with other  users, while fronthaul channels are transmitted through WDM point-to-point channels \cite{COMBO-JOCN}, in order to meet the strict latency requirements of fronthaul.
However using dedicated wavelengths for fronthaul is not always a cost effective solution, especially when the remote radio head is constituted by a small cell that does not require a 10G channel capacity. The 5G-PPP project 5G-Crosshaul \cite{Crosshaul-paper} is looking at architectures for transmission of both fronthaul and backhaul traffic over a common packet switched network. The development targets a unified  data plane protocol and switch architecture that meet the strict latency and jitter demands. From a standardisation perspective, as previously mentioned, the IEEE is currently working on similar topics with the P1914.1 Standard for Packet-based Fronthaul Transport Networks targeting architecture and technology to allow the transport of next generation fronthaul systems over Ethernet.
Another 5G-PPP project working on fixed/mobile convergence is the 5G-Xhaul \cite{xhaul-wiley}, which enhances the convergence by introducing point-to-multipoint millimeter-Wave systems to provide wireless fronthaul to remote radio head from a macro-cell location. The macro cell is then connected to the BBU pool through a CPRI interface across the metro network. The optical network transport is carried out through a Time-Shared Optical Network (TSON) system capable of providing sub-wavelength switching granularity to optimise resource usage.

These examples show how the integration of fixed and mobile networks has become a critical factor affecting the success of next generation mobile services. In addition, as NFV is progressively moving network functions from dedicated telecommunications vendor equipments to virtual machines, including data centres into the big picture becomes important. The next section provides additional insights on the integration of data centres in the network convergence.

Before concluding this section, we would like to focus the reader's attention to the trade-off between node consolidation, which is based on the use of longer access reach technologies, and services like fronthaul, which, due to their tight latency constraint, require a shorter optical reach. We believe that 5G network design should take both aspects into consideration, introducing in the ODN the flexibility to redirect latency-bound signals to local processing nodes, while allowing other signals to benefit from the network equipment consolidation enabled by the use of longer-reach connections. However, this trade-off has not been yet properly investigated and will require further study, which should be addressed by future research projects.

\subsection{Data centre integration}
\begin{figure*}
    \centering
    \includegraphics[width=0.8\textwidth]{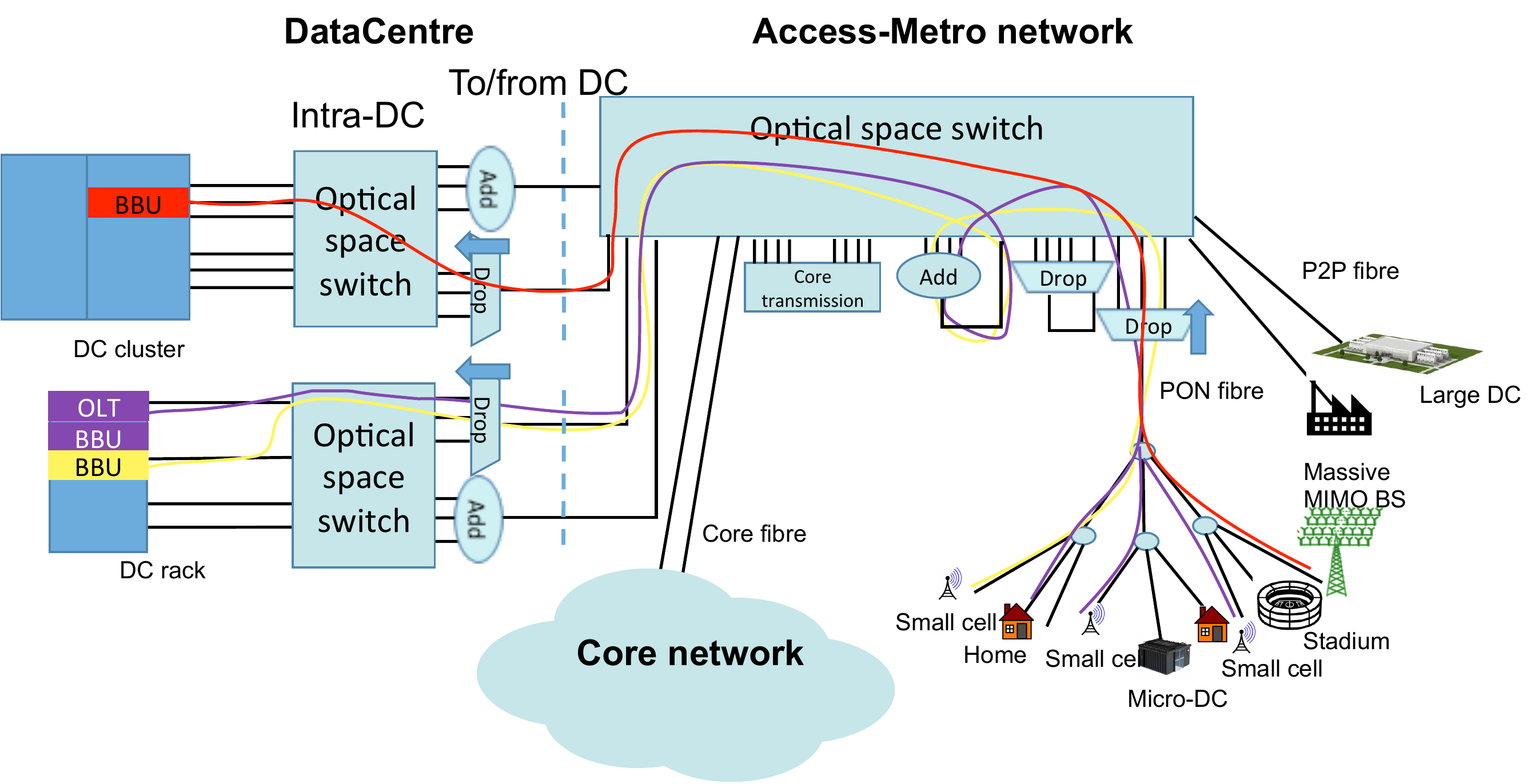}
    \caption{Integration of the data centre into the converged access-metro architecture through agile optical switching}
    \label{fig:DC-metro-arch}
\end{figure*}

Considering that much of the 5G framework revolves around virtualisation of networks and functions, integration of access-metro and data centre networks becomes essential for delivering the end-to-end vision. Taking as example a cloud RAN system, where the processing chain goes from the radio interface to the server where the networking processing functions are carried out, strict latency requirements for the service can only be met if the network and processing resources can be controlled across the entire chain \cite{Comms-standard-virtualisation-5G}.
As NFV moves network processing towards general purpose servers, effective scale up of processing power will require a re-design of the central office architecture, which will progressively migrate towards a typical data centre network, a view shared by the Central Office Re-architected as a Datacentre (CORD) \cite{CORD} collaborative project between AT\&T and ON.Lab. 
Integration of virtual resources from the wireless and optical domains to the DC was also one of the main research themes of the European CONTENT project, which introduced the idea of using the above mentioned TSON in the metro node for convergence of wireless, optical and DC resources. The CONTENT approach shows that it is possible to satisfy future content distribution without the use of cloudlets\footnote{A cloudlet is a small data centre located to the edge of the Internet, close to the end user, to provide computation resources with lower reduce latency.}, thus reducing overall energy consumption \cite{CONTENT-globecom} and increasing resource usage efficiency \cite{CONTENT-JOCN}, while only paying a small penalty in overall latency. It should be noticed that these studies predate the 5G-Xhaul project and focused on LTE backhaul, thus latency requirements were not an issues as for fronthaul.
When considering X-haul of next generation 5G systems, as anticipated in figure \ref{fig:pon_multi-service}, it is expected that a 5G flexible architecture should be capable of offering data processing elements in different parts of the network, and assign them to the application depending on its latency and capacity requirement. It should be expected that some network functions will be processed near the access point, some in the ODN near a branching or remote node, some at the central office and others in larger data centres.

Due to the increasing role data centres are playing in the converged network vision, the optical networking community has dedicated substantial effort in exploring novel technologies and architectures for faster interconnection within the data centre (intra-DC) and among data centres (inter-DC). The former has received particular attention on projects targeting next generation Exascale computing systems, through network architectures that make increasingly use of optical switching technology.  This includes study of hybrid electronic-optical switching architectures \cite{Elios}, \cite{Kostas-JOCN}, and circuit switching technologies ranging from 3D Micro Electro Mechanical Systems (MEMs) and fibre steering, capable of 10-20 ms switching times, to faster 2D MEMs \cite{2D-MEMs} capable of few microseconds switching times, to the combination of fast laser wavelength tuning and Array Wave Guide (AWG) with nanosecond switching times.

As the intra-DC network achieves faster optical switching, it becomes interesting to investigate the use of this technology outside the data centre, not only for connections among DCs but between DCs and any other 5G infrastructure, which we dub \textbf{\textit{DC-to-5G communication}}. Work in \cite{Samadi-optics-express} has demonstrated the SDN-driven inter-DC switching of transparent end-to-end optical connections across multiple metro nodes, using an architecture similar to that shown in figure \ref{fig:DC-metro-arch}, with commercial Reconfigurable Add Drop Multiplexers (ROADMs) for routing wavelengths across the nodes. The authors showed an overall path setup time of about 88ms, mostly dominated (72 ms) by the reconfiguration time of the ROADMs and optical switch, showing the potential for fast inter-DC optical circuit switching for applications such as virtual machine migration.

The ultimate vision is that of an agile access-metro optical network capable of transparently interconnecting end points, for example a RRH to the server or rack in the DC where the BBU is located, avoiding any intermediate packet processing, thus eliminating any further transport delay.  
A use case is shown in figure \ref{fig:DC-metro-arch}, operating over an Architecture on Demand type of node \cite{Bristol-AoD-JOCN}, where an optical space switch is used to enable dynamic reconfiguration of optical links, where wavelengths can be routed to the desired destination, at an access PON, at a DC, or else towards the network core (after undergoing any required aggregation into larger data streams, protocol conversion or other type of signal processing). The use case merges the concepts of C-RAN and NFV through the virtualisation of the PON Optical Line Terminal (OLT). It considers a number of small cells in a given area using mid-haul as transport mechanism (traffic is only shown in the upstream direction for simplicity): the traffic they generate on the transport link is three to six times higher than the traffic in the radio access side and is proportional to it \footnote{Considering that fronthaul can increase the data rate up to about 30 times that of the radio interface and that mid-haul can reduce this value by 5 to 10 times according to  \cite{Pfeiffer-JOCN},\cite{Miyamoto-split-PHY-OFC}, we have that the midhaul rate will be just three to six times that of the radio interface}.
When the traffic from those cells is too low to justify the use of dedicated wavelengths for each cell, their mid-haul traffic is aggregated over a TDM PON. The idea is to operate transparent switching of the PON wavelength channel at the access-metro node, terminating it directly into a rack or server in the DC (the violet link in figure \ref{fig:DC-metro-arch}), which implements both OLT\footnote{Notice that the MAC of the OLT might be operated in a merchant silicon hardware switch, as in the CORD architecture} and BBU (purple line terminating on purple OLT and BBU boxes). The colocation of OLT and BBU will minimise end-to-end latency, allowing for example synchronisation of OLT and BBU scheduling mechanisms or advanced PHY functions such as CoMP and coordinated beamforming among the BBUs.
A sudden increase in the small cells traffic however will likely create congestion in the PON, as for example a cell offering a peak rate of 1 Gbps could increase the load in the PON by 3 Gbps. The associated mid-haul traffic should thus be quickly allocated more capacity, by moving the transmission wavelength of the RRH to a dedicated channel (e.g., a 10G point-to-point channel), which gets transparently switched in the access-metro node towards the same BBU server (yellow line terminating on yellow BBU box), or if needed to a different server or even DC (red line), although this would require ultra-fast Virtual Machine migration. We believe the switching time should be at least comparable with that of the LTE scheduling time, i.e., of  the order of the millisecond, in order to achieve seamless transition. This requires the optical switches in the access-metro node and the DC to operate in the tens to hundreds of microseconds, suggesting the use of 2D-MEMs type of devices, and the implementation of an agile control plane capable of supporting this fast switching (switching control times of 10 microseconds were demonstrated for intra-DC environments in \cite{Porter-Sigcomm}).

Another issue worth mentioning is that mid-haul systems, similarly to fronthaul ones, require low transport latency (few hundreds microseconds), which cannot be achieved with Dynamic Bandwidth Assignment (DBA) mechanism available it today's PONs (operating on the order of few milliseconds). However studies in \cite{OFC-fix-mob-scheduling} have shown the possibility to reduce such values to few tens of microseconds by using DBA algorithms synchronised with the BBU scheduler.

\begin{figure}[h]
    \centering
    \includegraphics[width=\columnwidth]{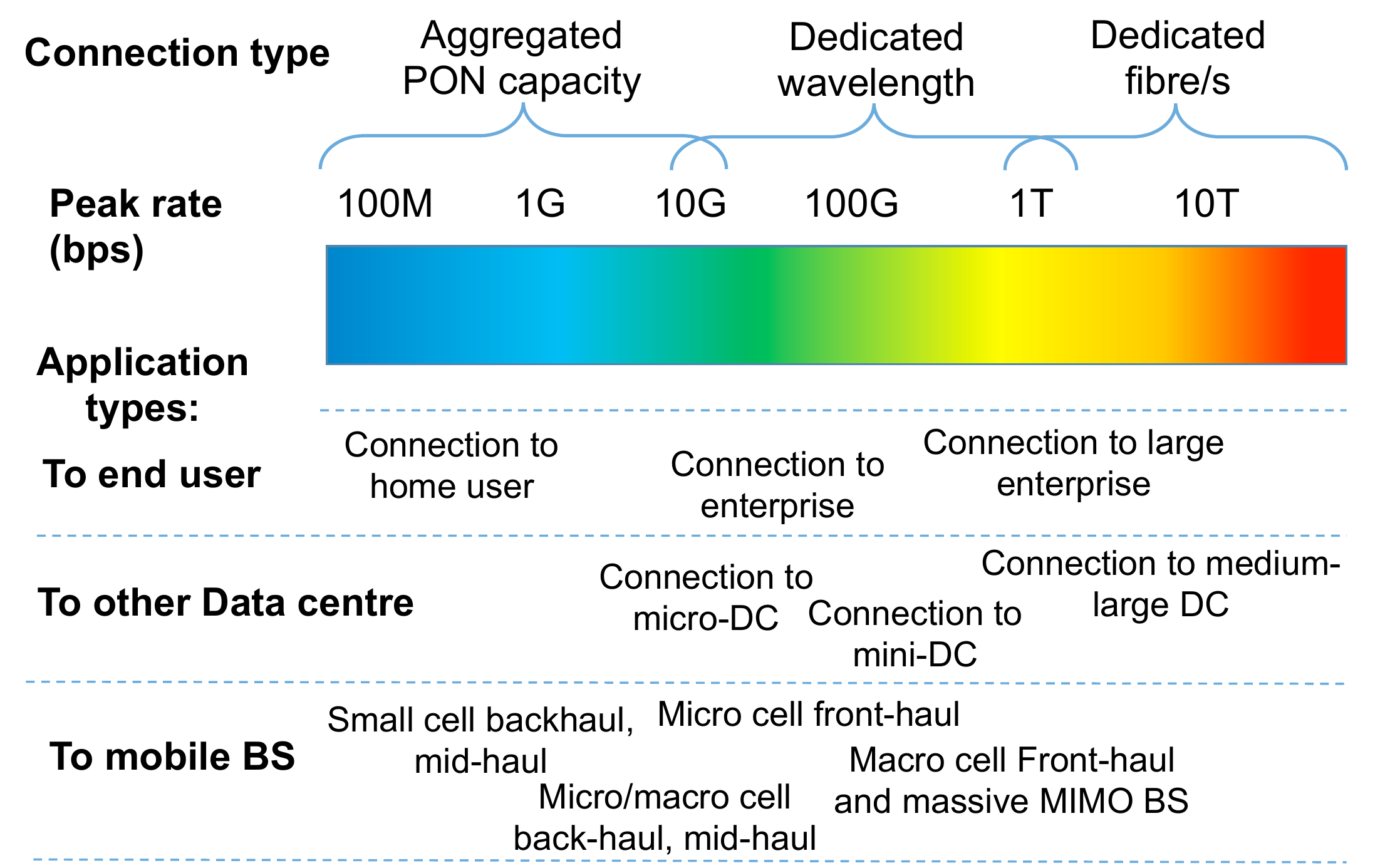}
    \caption{Example of service mapping from data centre to access network users.}
    \label{fig:conn-type}
\end{figure}

Finally, while this section has focused on a C-RAN case study, the concept of access-metro and agile data centre convergence can support other scenarios. Figure \ref{fig:conn-type} provides further examples, showing the potential mapping between services and transmission channels (i.e., shared TWDM PON channels, dedicated wavelength channels and dedicated fibre links).  
In addition, while it is recognised that sub-millisecond switching might be still a few years away in access-metro nodes, in the shorter term the ability to provide dynamic optical switching at lower speed will still prove useful to operate dynamic resource allocation at the network management level.

\  \

In conclusion, this section has shown how the seamless integration of wireless, optical and data centre networking technologies will be instrumental to provide the necessary network agility to satisfy dynamic end-to-end allocation of connections with assured data rate and latency requirements. The increasing use of transparent optical switching in the metro architecture will increase the ability to adopt new technologies as they become available. In addition, the increasing use of virtualisation to provide independency among network resource slices and end-to-end control of the leased resources will pave the way to application-oriented, and multi-tenant solutions, further described in the next section.

\section{Convergence in the ownership dimension: multi-tenancy}
\begin{figure}[h]
    \centering
    \includegraphics[width=\columnwidth]{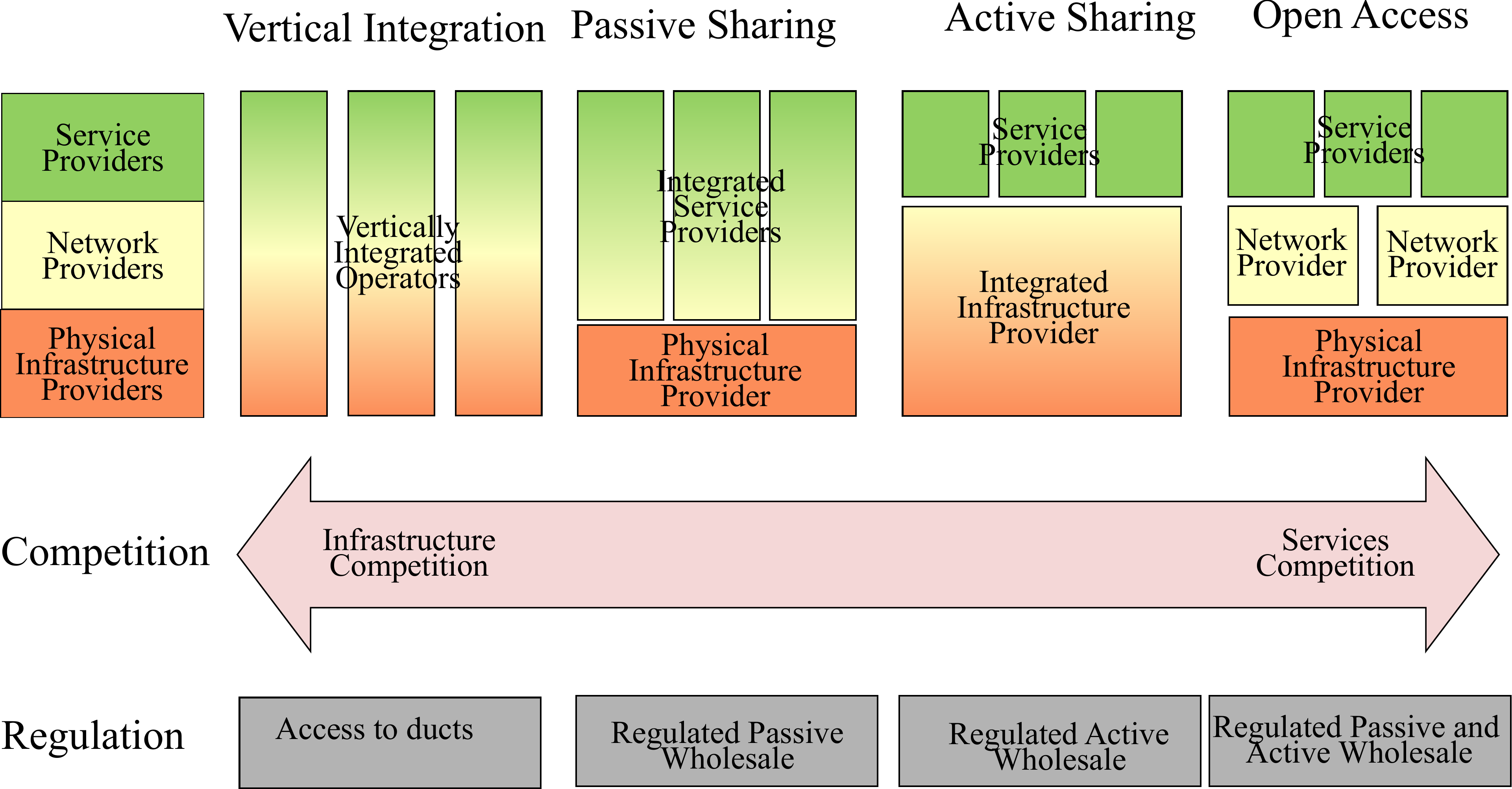}
    \caption{FTTH business models \cite{FTTH_council}}
    \label{fig:open-access-models}
\end{figure}

The last aspect of the convergence we want to discuss in this paper is in the ownership dimension, i.e. the ability of the network to support multi-tenancy. 
A simplified though common division of network ownership domains is in three layers:
\begin{itemize}
\item the passive network infrastructure: ducts, sub-ducts, fibre with any passive splitter element, and copper on some legacy networks; 
\item the active network infrastructure: transmission and switching equipment; 
\item the service layer: the provisioning of services to the end user. 
\end{itemize}
Multi-tenancy in access and metro networks has been extensively studied in the literature \cite{open-access-FTTH}-\cite{DISCUS-WP2} and figure \ref{fig:open-access-models} \cite{FTTH_council} shows different possible ownership models, ranging from total vertical integration, typical of incumbent operators that own all three layers, to complete separation where each layer is owned by a different entity.

While it is out of the scope of this paper to further investigate the pros and cons of these models, we argue that open access, at least of the active network infrastructure, is required in order to share infrastructure costs and open up the market to better competition. Access network sharing has been implemented in the past, though at a basic level, as  telecommunications regulators have enforced Local Loop Unbundling, (operating at the level of the passive infrastructure) and bitstream services \cite{DIW-Berlin} (at the active infrastructure level). Active network sharing has gained popularity as it allows Service Providers (SPs) to sell bandwidth services without owning physical network infrastructure, and the bitstream service has further evolved to Virtual Local Loop Unbundling (VULA) and next generation access (NGA) bitstream \cite{ALU-NGA}. However, although they give service providers more control, by adding some quality of service differentiation, they do not offer the ability to fully customise services to provide highly differentiated products to the users, and rely on the infrastructure owner for functions such as performance and fault monitoring, which are essential for serving the business sector. 

Work is currently underway by standardisation bodies such as the Broadband Forum under the Fixed Access Network Sharing study group \cite{BBF-SD-351} to increase the control of SPs over the access network through virtualisation. The vision is that of an ownership model differentiating between \cite{Cornaglia-OFT}:
\begin{itemize}
\item an Infrastructure provider (IP) that owns and maintains physical networking infrastructure (for example the passive and active network infrastructure), enables physical resource virtualisation and carries out the network virtualisation, provides virtual resources controlling APIs for the Virtual Network Operator (VNO) and gets its revenue by leasing resources to the VNO.
\item a Virtual Network Operator that operates, controls and manages its assigned virtual network instance, is able to run and re-designs customised protocols in its own virtual networks, provides specific and customised services through its own virtual networks, receives revenue from the end users and pays the usage of network resources to the IP, while saving on network infrastructure deployment costs.
\end{itemize}

The proposed roadmap is through three sequential steps, as suggested in \cite{Cornaglia-ECOC}:
\begin{enumerate}
\item the first step is to reuse existing network equipment controlled by the infrastructure provider through their management system: virtual network operators could get access to the network through a standard interface which can provide raw access to the management layer with optional monitoring and diagnostic functionalities. 
\item the second step is to deploy new hardware in the access node capable of resource virtualisation, so that the virtual network operators could be assigned a virtual network slice and get full control of the equipment. For this step however the interface remains, similarly to step 1), to the network operator management system. 
\item the third and final step is the full SDN integration, where the virtual operators access their network slice through a flexible SDN framework with standardised APIs. 
\end{enumerate}
\ \

In conclusion, multi-tenancy enables network infrastructure sharing, reducing the cost of ownerships and opening the market to a plurality of new entities that can provide the diversity that is necessary to empower the 5G vision.

\begin{figure}[h]
    \centering
    \includegraphics[width=\columnwidth]{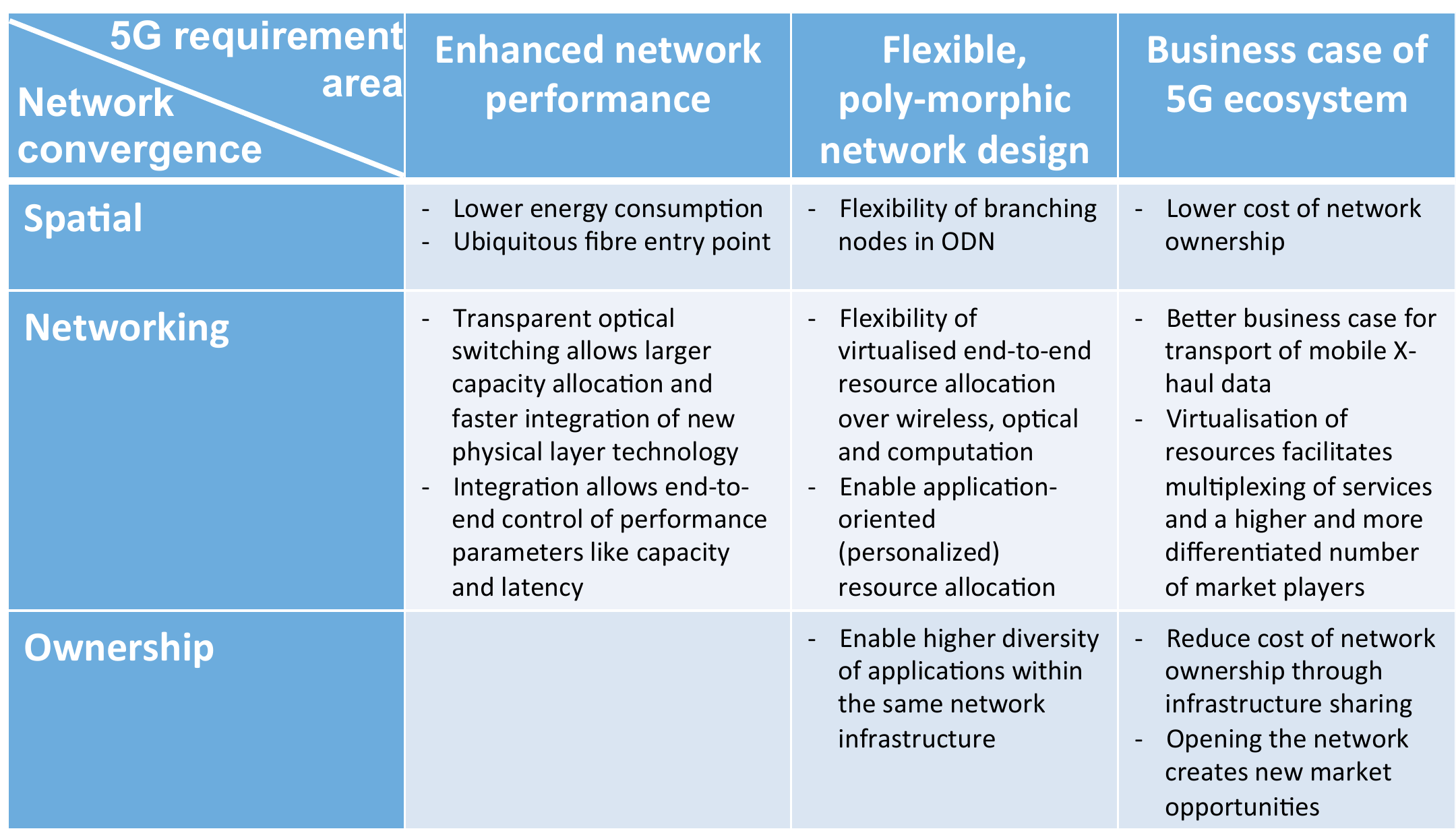}
    \caption{Summary of mapping between network convergence domain and 5G requirements}
    \label{fig:req_mapping}
\end{figure}

Although we do not suggest that the three dimensions of convergence described in this paper constitute a complete set for satisfying all foreseeable requirements of 5G networks, we argue that they can contribute to its realisation.
A summary of the mapping between the described convergence dimensions and the 5G requirements, grouped into three activity areas as outlined in section IV, is reported in figure \ref{fig:req_mapping}.

\section{Final remarks on the role of SDN and new business models}
After reporting on a number of research activities on network convergence and showing how they can contribute to meet future 5G requirements, we finalise the paper by briefly considering two further aspects. The first is a contemplation of the role that SDN could play in this future convergence and the second is a vision on future end-to-end virtualisation and associated business models.

We believe SDN will play a pivotal role in the multi-dimensional convergence discussed above, both as the mechanism to orchestrate the interaction among the different network domains and technologies, and the system to integrate and automate many of the control and management operations that are today still carried out through proprietary and closed interfaces. While SDN has experienced a hype in the past few years (which has also caused disagreement over its capabilities) there is now firm interest from industry, with concepts moving from research to production environment \cite{ATT-domain}, and with many related working groups being established in different standardisation bodies, including ETSI, ONF, BBF and IEEE. The number of enterprises, including many startups, working profitably in the SDN industry is constantly increasing, with an estimated overall value of \$15 billion in 2015, and a projected Compound Annual Growth Rate (CAGR) of 46\% for the next 5 years, bringing the market value to \$105 billion by 2020 \cite{SDX-central-report}.

Besides reducing cost of network ownership, by making more efficient use of the network capacity, reducing energy consumption, and enabling the convergence of multiple services (business and residential) into the same physical infrastructure, SDN will generate new form of revenues, by allowing operators to quickly deploy new services in their existing infrastructure.

One of these will come from enabling end-to-end virtualisation of the network, which will provide the opportunity for new business models for network sharing. This process, started in the U.S. with the physical unbundling enforced by the telecom deregulation act in 1996 \cite{FCC-deregulation}, is now evolving to virtualised bitstream concepts offering higher flexibility and product differentiation, and will continue to evolve towards full end-to-end network virtualisation \cite{OSHARE}.
The idea is to break down the network into a pool of virtual resources whose lease is negotiated to provide dedicated end-to-end connections, across multiple network domains and technologies. For example a customer could submit a service request through a user portal that is forwarded to a network orchestrator acting as a resource broker. A digital auction is then carried out to secure a chain of virtual resources to provide an ephemeral end-to-end connection \cite{EuCNC-Johann} that only lasts for the time required by the service, releasing all resources immediately after the service terminates. In principle this could be carried out with high granularity targeting network connection performances of individual flows, although more work is required to determine its scalability. 
\  \

The practical implementation of such concept will likely require a shift towards business models that better reflect the value chain of the service offered. From an end user perspective, the value is in the service delivered and its expected performance. Acceptable performance values such as data rate, latency and packet loss vary among services and thus need to be assured on a service base and not in average across all services delivered to an end point. For example the value per bit of a medical application monitoring and transmitting vital body parameters to a medical centre will be much higher than that of a video streaming service. It thus makes sense for the end user to pay for the service rather than for the network connectivity. The network should be paid by the entity providing the service which would select the most appropriate quality of service that assures reliable service delivery (thus paying different price levels for personalised ephemeral connections as mentioned above). This will make network operations completely transparent to the end users, which only interacts with the services and application of interest. Some basic forms of this idea have already been implemented by industry. For example, users of the Amazon kindle 3G pay for the purchase of their books but not for the network delivering it; Facebook zero and Google free zone are similar examples where the user is able to access a small set of their services without paying for the underlying network.  

Even if an agreement on potential business models is still far ahead, the idea of providing the option of better quality of service for targeted applications is gathering interest among network operators \cite{BBF-SD-377}, as many believe SDN can now provide the level of programmability and network automation required to practically implement this idea on a large scale.

Finally, while it is still an open question whether SDN will be capable of enabling the full network convergence and end-to-end virtualisation envisaged in future 5G networks, we would like to justify this optimism by pointing out the successes it has already enabled. A pragmatic analysis reveals that SDN has so far boosted research in several areas of networking, by producing a framework where testing of new algorithms, protocols and architectures \cite{ONDM-Flatland}, can be quickly moved from simulation/emulation environments (e.g., using the mininet platform \cite{mininet}) to real testbed scenarios \cite{OFT-paper}, \cite{1-N-protection}, enabling large-scale interoperability tests among different research groups and physical laboratories \cite{EuCNC-DISCUS-IDEALIST}, \cite{ONDM-DISCUS-IDEALIST}.
 
\section{Conclusions and future work}
The main contribution of this paper was to present a novel perspective on network convergence, proposing a multi-dimensional, end-to-end approach to network design to enable future 5G services. 
While providing a tutorial of the technologies, architectures and concepts revolving around network convergence, we have presented a number of high level concepts that we believe are of paramount importance in the 5G vision. 
One of the key messages is that an un-converged view of the network based on the piece-wise development of the individual segments is suboptimal and can lead to uneconomical decisions. A typical example is that viewing the wireless domain only as a client of the optical domain has lead to issues in performance and deployment cost. 

Network architectures should be designed with an end-to-end vision in mind, as the real value for the end user comes only from a consistent support of the application from its source to its destination. This is becoming increasingly clear among the networking community, as for example the European Commission is promoting an increasing number of ICT funding calls targeting converged fixed and wireless architectures towards 5G.

Finally, we have identified many open areas of research, which we believe will become increasingly relevant in the coming years. We conclude the paper outlining a few:
\begin{itemize}
\item aggregation of mobile mid-haul links over TWDM PONs; 
\item end-to-end optimisation of dynamic resource allocation, including wireless spectrum and antenna resources, optical transport and processing resources in the data centre;
\item architectural solutions capable of merging the benefits of node consolidations with latency-bound requirements of some of the 5G services and applications;
\item joint development of protocols and scheduling algorithms across the fixed and wireless domains;
\item convergence of access-metro and DC through faster optical switching enabling DC-to-5G communication;
\item scalable, SDN-controlled, end-to-end QoS assurance;
\item new application-centric business models.
\end{itemize}

\section*{Acknowledgment}
The author would like to thank Prof. David B. Payne for the fruitful discussions on the Long-Reach Passive Optical Network architecture and Prof. Linda Doyle for the insightful conversations over 5G networks and services.

\ifCLASSOPTIONcaptionsoff
  \newpage
\fi

\begin{IEEEbiography}[{\includegraphics[width=1in,height=1.25in,clip,keepaspectratio]{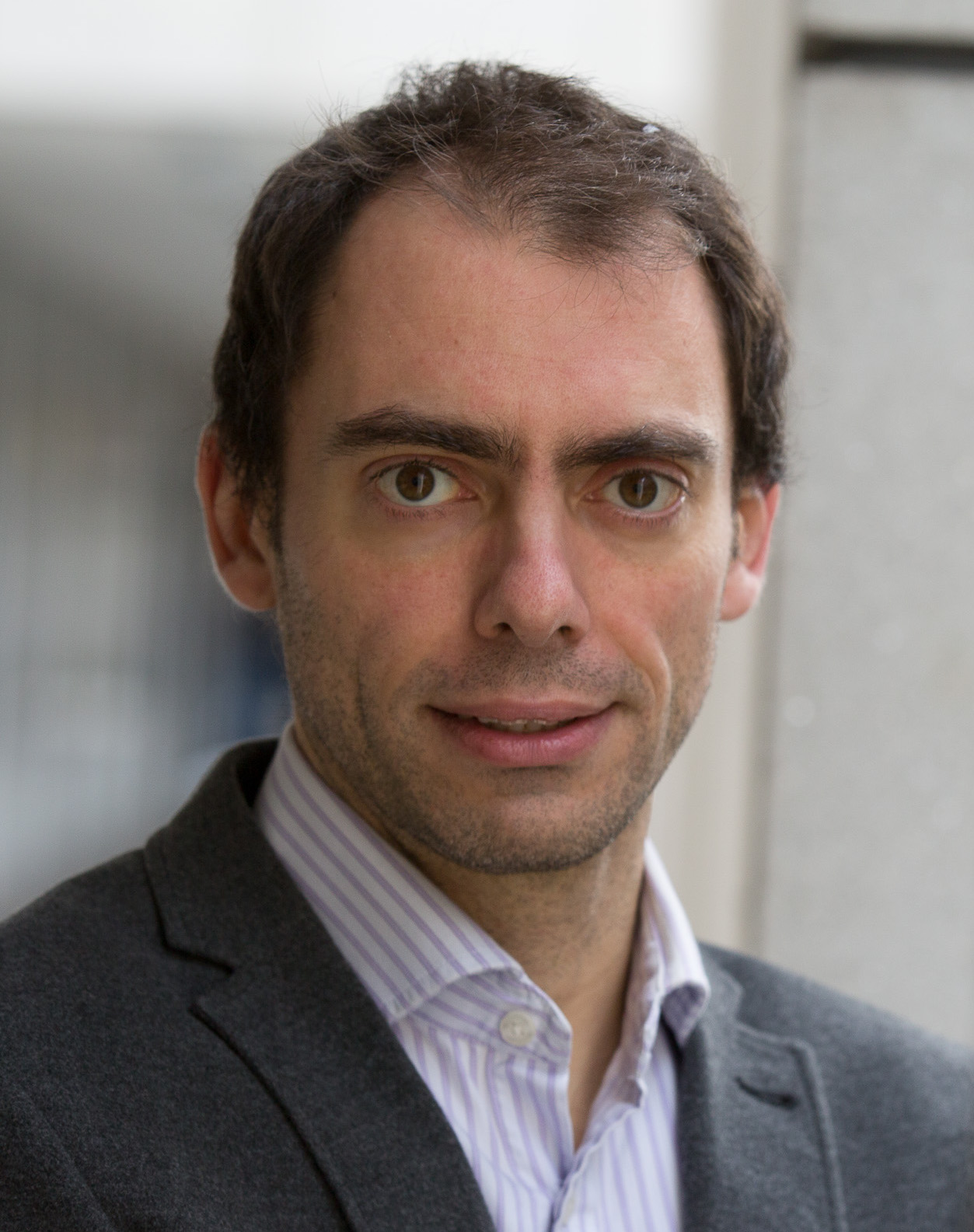}}]{Marco Ruffini}
received his M.Eng. in telecommunications in 2002 from Polytechnic University of Marche, Italy. After working as a research scientist for Philips in Germany, he joined Trinity College Dublin (TCD) in 2005, where he received his Ph.D. in 2007. Since 2010, he has been Assistant Professor (tenured 2014) at TCD. He is Principal Investigator at the CTVR/CONNECT Telecommunications Research Centre at TCD, currently involved in several Science Foundation Ireland (SFI) and H2020 projects, and leads the Optical Network Architecture group at Trinity College Dublin. He is author of more than 80 journal and conference publications and more than 10 patents. His research focuses on flexible and shared high-capacity fibre broadband architectures and protocols, network convergence and Software Defined Networks control planes.

\end{IEEEbiography}


%\newpage


%\vfill


\end{document}